\documentclass[conference,compsoc]{IEEEtran}

\ifCLASSOPTIONcompsoc
  \usepackage[nocompress]{cite}
\else
  \usepackage{cite}
\fi

\usepackage{graphicx}
\graphicspath{{figures/}}

\usepackage{tikz}
\usetikzlibrary{arrows.meta,positioning,fit,backgrounds,calc}

\usepackage{amsmath}
\usepackage{amssymb}
\usepackage{url}
\usepackage{xcolor}
\usepackage{booktabs}
\usepackage{multirow}
\usepackage{listings}

\usepackage{amsthm}
\usepackage{algorithm}
\usepackage{algpseudocode}
\theoremstyle{definition}
\newtheorem{definition}{Definition}
\newtheorem{property}{Property}

\newcounter{myidx}

\newcommand{\nextidx}{%
  \refstepcounter{myidx}%
  \arabic{myidx}%
}

\newcounter{myidx2}

\newcommand{\nextidxU}{%
  \refstepcounter{myidx2}%
  \arabic{myidx2}%
}

\newcommand{\predcat}{\par\medskip\noindent}

\newcommand{\tx}{\tau}
\newcommand{\txr}{r}

\lstdefinelanguage{Solidity}{
  keywords={
    pragma,contract,abstract,interface,library,function,returns,return,
    external,public,internal,private,view,pure,payable,virtual,override,
    if,else,revert,address,uint256,bytes,bytes4,calldata,memory,storage
  },
  keywordstyle=\color{blue!70!black}\bfseries,
  sensitive=true,
  comment=[l]{//},
  commentstyle=\color{green!40!black},
  morestring=[b]",
  stringstyle=\color{red!60!black},
}

\lstdefinestyle{soliditysnippet}{
  language=Solidity,
  basicstyle=\ttfamily\footnotesize,
  columns=fullflexible,
  keepspaces=true,
  showstringspaces=false,
  frame=single,
  framerule=0.3pt,
  rulecolor=\color{black!20},
  breaklines=true,
  aboveskip=0pt,
  belowskip=0pt
}

\usepackage[hidelinks]{hyperref}
\hypersetup{
  pdftitle={Ghost-Filled Orders: Detecting and Testing Atomicity Violations in Non-Custodial Prediction Markets},
  pdfauthor={Zhiyang Chen, Fan Long, Zhendong Su},
  pdfsubject={Atomicity violations in non-custodial prediction markets},
  pdfkeywords={prediction markets, blockchain, atomicity, ghost-filled orders}
}

\begin{document}

\title{Ghost-Filled Orders: 
Detecting and Testing Atomicity Violations \\
in Non-Custodial Prediction Markets}

\author{
\IEEEauthorblockN{Zhiyang Chen}
\IEEEauthorblockA{University of Toronto \& ETH Zurich\\
\texttt{zhiychen@cs.toronto.edu}}
\and
\IEEEauthorblockN{Fan Long}
\IEEEauthorblockA{University of Toronto\\
\texttt{fanl@cs.toronto.edu}}
\and
\IEEEauthorblockN{Zhendong Su}
\IEEEauthorblockA{ETH Zurich\\
\texttt{zhendong.su@inf.ethz.ch}}
}

\maketitle

\begin{abstract}
Blockchain based prediction markets combine offchain order 
management with onchain settlement. This architecture 
supports user controlled custody, since users keep funds in 
wallets or smart contracts while submitting signed orders 
to an offchain order book. 
However, it creates an atomicity gap. An order may be valid 
when accepted or matched offchain, but become invalid 
before the corresponding onchain settlement transaction 
is executed. This behavior, often called ghost filled 
orders by the community, can cause trades that appear 
filled offchain to fail onchain.
This paper studies this atomicity gap through a case 
study of Polymarket. We show how the delay between 
offchain order acceptance and onchain settlement allows 
adversaries to invalidate unfavorable orders after 
observing market outcomes or price movements. 
We then quantify the scale and financial impact 
of this behavior over a nine-month period from August 12, 2025,
to May 22, 2026, using 1.8 million reverted transactions 
involving Polymarket official smart contracts. Our analysis 
separates attacker profit from market and user loss, and 
develops conservative measurement rules to avoid overclaiming impact.
We further develop a methodology for testing other 
blockchain based prediction markets and apply it to three 
additional markets. We find that all three are 
vulnerable to the same class of attack, and that 
one design enables direct attacker profit. We have 
reported the findings to all three projects; one project
had acknowledged the issue at the time of the study.

\end{abstract}

\pagestyle{plain}
\thispagestyle{plain}

\section{Introduction}
\label{sec:intro}

Prediction markets are markets for contracts 
whose payoffs depend on future
event outcomes, and their prices are often 
interpreted as aggregated
probabilistic forecasts of those outcomes~\cite{wolfers2004prediction}.
The market price of a contract therefore provides a real-time
estimate of the probability assigned by market participants to the corresponding
outcome. Recent adoption has made prediction markets a 
significant part of the
financial system: as of June 5, 2026, it is reported that 96 prediction markets
have \$480.16M in total value locked and produce \$2.709B in seven-day
prediction volume~\cite{defillama_prediction_market}.

Prediction markets are commonly implemented in two forms: centralized markets,
such as Kalshi~\cite{kalshi}, and decentralized, blockchain-based markets, such
as Polymarket~\cite{polymarket}. 
In a centralized prediction market, the operator holds user deposits 
and records trades in an internal ledger. 
In a decentralized market, users retain control of their funds, 
submit orders to an \textit{off-chain} relay, and rely on smart contracts 
to settle matched trades \textit{on-chain}. 
Compared with centralized markets, the blockchain-based
design is appealing for security and transparency: 
users keep custody of their own assets, 
auditors can inspect settled orders and on-chain smart contracts, 
and privacy-conscious traders can participate without revealing
their real-world identities.

However, the same hybrid architecture that makes non-custodial prediction markets
practical also creates a new security problem: atomicity violations. 
To avoid the cost and latency of recording every order-book update 
on-chain, these decentralized prediction markets accept and match 
user-signed orders off-chain and defer settlement to an on-chain exchange contract.
As a result, an order can be valid when accepted and matched off-chain, 
yet become invalid by the time settlement occurs on-chain. 
This behavior stems from a classic time-of-check to time-of-use (TOCTOU) 
gap~\cite{bishop1996checking}: the operator evaluates 
an order against one blockchain state, while the exchange contract 
enforces settlement against a later state.
This gap is exploitable because the on-chain preconditions 
of a signed order remain mutable. 
The order maker can still revoke allowance, move the tokens 
backing the order, or otherwise alter the blockchain
state required for settlement. 
If this invalidation happens after the order has matched and
moved from the off-chain order book, the 
order becomes a \emph{ghost order}: 
visible to other traders and already shaping the displayed price on 
front-end interfaces, 
yet no longer fillable on-chain. 
When timed around fast-moving information, 
such as the end of a sports game, 
ghost orders effectively grant their makers a 
unilateral option: they can let favorable orders settle while 
invalidating unfavorable ones.

This threat is not hypothetical. In 2026, Polymarket publicly acknowledged 
the issue of ``ghost-filled'' orders~\cite{polymarket_ghost_ack1,
polymarket_ghost_ack2}, 
and subsequently announced two urgent  
upgrades~\cite{polymarket_fix1,polymarket_fix2}. 
These events raise several
questions unanswered that are important for Polymarket
users, other prediction market users and security researchers. 
\begin{itemize}
  \item What actions can invalidate a previously matched order and cause on-chain 
    settlement to revert?
  \item How can malicious ghost-filled orders be 
    distinguished from benign settlement failures and operational mistakes?
  \item How prevalent are ghost-filled orders in practice, and what financial impact 
    do they impose on traders, market makers, and the market as a whole? 
  \item Is this problem unique to Polymarket, or do other non-custodial
    prediction markets expose similar atomicity violation vulnerabilities?
\end{itemize}

This paper presents a systematic study of ghost-filled orders in
non-custodial prediction markets. We formulate the problem as an off-chain/on-chain
atomicity violation: 
an adversary exploits the delay between off-chain
order fill and on-chain settlement by invalidating 
the blockchain state required for settlement.
We then study this issue in Polymarket using $1.8$ million
reverted transactions involving  Polymarket smart contracts over a nine-month
period from August 12, 2025 to May 22, 2026. Our analysis reconstructs the failed
settlement attempts, classifies the violated preconditions, backtracks
the order-invalidating transactions, and identifies the responsible parties.
Our study yields three main findings. 
First, ghost-filled orders are not
uniform in nature. A substantial share exhibit
patterns that indicate deliberate, strategic 
invalidation: they concentrate in a
small number of markets and responsible parties, 
and invalidate the settlement transaction 
within very short timeframes (0-2 blocks). 
Second, these reverts of orders have strongly asymmetric
economic effects: in timing-sensitive markets, 
reverted fills overwhelmingly
avoid losses for the order owner, 
consistent with strategic invalidation of
unfavorable trades. 
Third, the
underlying atomicity gap is not unique to Polymarket.
Applying our auditing procedure to three additional blockchain-based prediction
markets, we find that all three are exploitable through the same class of
invalidation, and one is directly profitable for potential attackers. 

\noindent\textbf{Contributions.} We make the following contributions:
\begin{itemize}
  \item \textbf{Problem formulation.}
  We formalize ghost-filled orders 
  as an atomicity violation problem in 
  non-custodial blockchain-based prediction markets.  
  We propose two 
  important research problems in this context: 
  analyzing existing ghost-filled orders, and
  auditing new prediction markets.

  \item \textbf{Analysis Framework.}
  We build a framework of \emph{detectors} 
  and \emph{probes} that classifies
  each reverted settlement by the violated
  precondition and backtracks their order-invalidating
  transactions. 
  Over all 1,824,926 Polymarket reverts, it
  classifies 100\% of reverts and
  identifies order-invalidating transactions 
  for  96.6\% of the user-controlled reverts.

  \item \textbf{Financial Impact Analysis.}
  Using our analysis framework, we evaluate two highly suspicious scenarios
  in which ghost-filled orders were very likely used to avoid losses or
  outcompete other market makers for potential profit.
  Our results show the responsible parties have 
  avoided a loss over 61M USD, and 77\% of the ghost-filled
  order clusters are dominated by a single responsible party.

  \item \textbf{Auditing Procedure and Zero-Day Discoveries.}
  We develop a black-box audit procedure and
  apply it to three other
  blockchain-based prediction markets: 
  surprisingly all three are still exploitable 
  and one is directly profitable. 
  We responsibly disclosed every finding, 
  and one market
  has acknowledged the vulnerability.

\end{itemize}

The rest of the paper is organized as follows.
Section~\ref{sec:background} provides the necessary background knowledge for
the rest of the paper.
Section~\ref{sec:ghost-filled-orders} describes the ghost-filled orders
and explains the mechanism behind them.
Section~\ref{sec:formulation} formalizes the 
definitions, threat model, and research problems that 
are explored throughout the remainder of the paper.
Section~\ref{sec:analysis} presents our analysis framework for analyzing 
ghost-filled orders on Polymarket. 
It reports the coverage of our analysis framework and 
presents two highly suspicious case studies 
identified, together with their quantified financial impact.
Section~\ref{sec:auditing} presents our auditing and testing procedure, 
our results on other blockchain-based prediction markets, and the 
possible defenses and mitigations.
Finally, Section~\ref{sec:related} discusses related work, and
Section~\ref{sec:conclusion} concludes the paper.

\section{Background}
\label{sec:background}

This section provides necessary background knowledge for 
understanding the rest of the paper.

\subsection{Prediction Markets}
\label{subsec:bg-prediction}
Prediction markets are exchanges where participants trade 
contracts whose payoff
is tied to the outcome of a future event, 
so that the market price of a contract
reflects the crowd's aggregate estimate of how likely 
that event is to
occur~\cite{wolfers2004prediction}. 
Prediction markets fall into two broad categories. 
\emph{Centralized} (custodial)
markets, such as Kalshi~\cite{kalshi}, take custody of user 
funds and run the
order book, matching, and settlement on their own infrastructure, 
so that users
trust the operator to hold balances and pay out correctly.
\emph{Decentralized} (non-custodial) blockchain-based markets, 
such as
Polymarket~\cite{polymarket},
instead settle
trades through on-chain smart contracts, so that users retain 
control over their
funds until settlement and need not trust an operator with custody.
In the rest of the paper, we use ``prediction market'' to refer 
to non-custodial
blockchain-based prediction markets, unless otherwise specified.

\subsection{Collateral and Conditional Tokens}
\label{subsec:bg-tokens}

\noindent \textbf{Collateral Tokens in ERC-20~\cite{erc20}.}
Polymarket uses stablecoins as collateral tokens: 
USDC.e~\cite{usdce} in Polymarket v1 and pUSD~\cite{pusd} in 
Polymarket v2. 
Both are ERC-20 tokens~\cite{erc20}. 
Collateral tokens are used to purchase outcome
shares, which are represented as conditional tokens.

\noindent \textbf{Conditional Tokens in ERC-1155~\cite{erc1155}.}
In a binary prediction market, each possible 
outcome is represented by an outcome token, 
that pays out if that outcome occurs.
Polymarket represents these outcome 
tokens using the Conditional 
Token Framework (CTF), which encodes each 
market outcome as an ERC-1155 
position identifier~\cite{polymarket_ctf,erc1155}. 
Each event therefore has two conditional tokens: Yes and No tokens.
Users can merge 1 Yes and 1 No token to get back 1 collateral token, 
or split 1 collateral token into 1 Yes and 1 No token.

\subsection{Order Book and Maker/Taker Orders}
\label{subsec:bg-orderbook}
Prediction markets commonly use a central 
limit order book (CLOB) to organize
trading interest. Orders resting in the book
are typically \emph{maker orders}, because 
they provide liquidity for future
matches. A \emph{taker} consumes this liquidity by 
matching against one or more
maker orders. 
In Polymarket-style settlement, both 
maker and taker orders may be buy orders or
sell orders. A buy order
pays collateral and receives 
conditional tokens, while a sell order pays
conditional tokens and receives collateral.

\section{Ghost-Filled Orders} 
\label{sec:ghost-filled-orders}

In a non-custodial blockchain-based prediction market such as 
Polymarket, order handling is split across 
four parties: the trader, the off-chain CLOB, the market operator, and the 
blockchain. A trader first signs an order and submits it to the CLOB. 
The CLOB checks the order and marks it as \texttt{LIVE} or 
\texttt{DELAYED}.\footnote{In some markets such as sports markets, 
Polymarket places marketable orders into an asynchronous 
seconds-delay window before matching them~\cite{polymarket_order_lifecycle}.} 
When a compatible taker order arrives, the CLOB matches it 
against one or more maker orders and 
marks the trade as \texttt{MATCHED}. Only after this 
off-chain matching step does a Polymarket-authorized operator 
submit a settlement transaction on-chain on behalf of the CLOB, which is supposed 
to execute the trade by transferring tokens between the matched parties and 
finalizing the market outcome.

\begin{figure}[!t]
\centering
\includegraphics[width=\columnwidth]{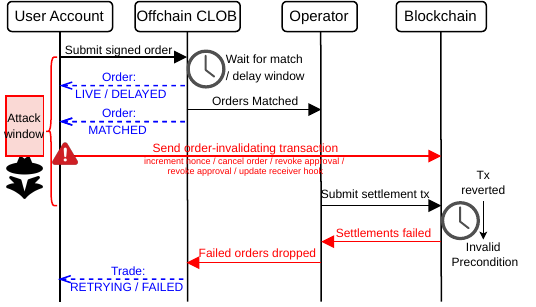}
\caption{High-level intuition for ghost-filled orders in a non-custodial
prediction market. Orders are accepted and matched off-chain, but settled
later on-chain. During this gap, the trader can still change nonce, balances,
approvals, or receiver configuration, so an order that appears matched in the
off-chain book may fail when the operator later submits the settlement
transaction.}
\label{fig:attack-flow}
\end{figure}

This lifecycle creates an atomicity gap between off-chain 
order submission and on-chain order
settlement. The gap includes the delay window, 
the time spent waiting for a compatible match, 
the time needed by the operator to construct and submit the settlement transaction, 
and the time needed for that transaction to be included and finalized on-chain. 
During this period, the order may appear valid and even matched in the CLOB, 
but the on-chain state required to execute the order is not locked.

Figure~\ref{fig:attack-flow} illustrates how this gap can be exploited. 
After an order is submitted or even matched off-chain, 
the trader can still submit a separate order-invalidating 
transaction to the blockchain. 
For example, the trader may increment the order nonce, 
move collateral or conditional tokens. 
These actions are ordinary user-controlled smart contract interactions,  
and do not require privileged access to the Polymarket contracts. 
When the operator later submits the settlement transaction, 
the Polymarket contracts  evaluate the order against the later 
blockchain state, not the earlier state observed by the CLOB. 
If any required precondition has been invalidated, the settlement 
transaction reverts. 
The off-chain system then mark the trade as 
\texttt{RETRYING} or \texttt{FAILED}, return the outcome to the CLOB, 
and drop the failed order. 
Such orders are called \emph{ghost-filled}: 
the order was accepted and matched off-chain, but the corresponding 
settlement fails on-chain, so the trade never actually executes. 

\noindent \textbf{Consequences of Ghost-Filled Orders.} The primary consequence is that the order owner obtains a unilateral
option over settlement. If new outcome-relevant information or price
movement arrives after the off-chain match but before on-chain settlement,
the owner can allow favorable fills to settle while invalidating unfavorable
fills. This shifts execution risk to counterparties whose matched trades
remain exposed to a later on-chain failure, and gives the invalidating
trader an advantage unavailable to ordinary users who expect matched orders
to be final.

A second consequence is that the impact can extend beyond the invalidated
order itself. 
A single settlement transaction may bundle multiple matched orders, 
and on-chain execution is atomic: 
if one invalidated order causes the transaction to revert, 
the entire settlement fails. 
Consequently, otherwise valid orders in the same bundle may 
also fail to settle. 
From the perspective of other users viewing the market through the 
frontend, a trade may appear to have been matched, only to later disappear 
or be reported as failed because the bundled on-chain settlement 
never completed. 

\section{Problem Formulation}
\label{sec:formulation}

This section outlines the preliminaries, threat model, 
and problem formulation for the rest of the paper.

\subsection{Preliminaries}
\label{subsec:preliminaries}

\begin{definition}[Transaction]
\label{def:tx}
A \emph{transaction} is a tuple
\[
  \tx=(c_{\tx},f_{\tx},d_{\tx},s_{\tx},v_{\tx},\mathit{gas}_{\tx},\sigma_{\tx}),
\]
where $c_{\tx}$ is the contract invoked by $\tx$, $f_{\tx}$ is the called function,
$d_{\tx}$ is the calldata, $s_{\tx}$ is the sender, $v_{\tx}$ is the transferred native
value, $\mathit{gas}_{\tx}$ is the gas limit, and $\sigma_{\tx}$ is the
transaction-level signature or authorization. 
\end{definition}
A reverted settlement transaction is denoted by $\txr$.

\begin{definition}[Order]
\label{def:order}
An \emph{order} is a tuple
\[
  \omega=(\mathit{owner}_{\omega},\mathit{role}_{\omega},\mathit{side}_{\omega},
          \iota_{\omega},p_{\omega},q_{\omega},n_{\omega},e_{\omega}),
\]
where $\mathit{owner}_{\omega}$ is the account that owns and signs the order,
$\mathit{role}_{\omega}\in\{\mathsf{maker},\mathsf{taker}\}$ is its role in the match,
$\mathit{side}_{\omega}\in\{\mathsf{buy},\mathsf{sell}\}$ is its side,
$\iota_{\omega}$ is the conditional (ERC-1155) token id being traded, 
$p_{\omega}$ is the price, $q_{\omega}$ is
the order amount, $n_{\omega}$ is the nonce, and $e_{\omega}$ is the expiry. 
\end{definition}

In Polymarket, an order is authorized by its owner through a signed order
payload and then submitted to the CLOB. 
The signed payload binds the order's
core on-chain settlement terms, including its asset, side, 
price, and amount.
Once submitted, an order is not modified in place: its owner cannot directly
edit fields such as the amount $q_{\omega}$, price, or effective expiry
$e_{\omega}$. To change these terms, the owner must cancel the resting order
and submit a new order payload, with a new signature where required.

\begin{definition}[Settlement bundle]
\label{def:bundle}
A settlement transaction $\tx$ bundles orders submitted together for
settlement. The \emph{settlement bundle} of $\tx$ is the ordered list of orders
\[
  \Omega(\tx)=(\omega_1,\ldots,\omega_K),
\]
where each $\omega_i$ is an order in the sense of Definition~\ref{def:order}, and the
orders are indexed in the sequence in which they are filled during execution, as
observed in the transaction trace.\footnote{The fill sequence can be recovered from the
  \texttt{OrderFilled} events emitted by the Polymarket exchange contracts from 
  the trace of replaying $\tx$; each event
  records the order hash and the filled quantity, and their emission order gives the
  sequence in which the orders are filled.}
\end{definition}

\begin{definition}[Blockchain state]
\label{def:blockchain-state}
A \emph{blockchain state} is a mapping from account addresses to account states, where
each account state comprises the account's balance, nonce, code, and storage. We write
$\mathcal{S}$ for the set of all blockchain states. For a transaction $\tx$, we write
$S_{\tx}\in\mathcal{S}$ for its \emph{prestate}, the state immediately before $\tx$ is
executed, and $S^{\prime}_{\tx}\in\mathcal{S}$ for its \emph{poststate}, the state
immediately after.
\end{definition}

\begin{definition}[Precondition predicate]
\label{def:precondition-predicate}
A \emph{precondition predicate} is a Boolean function
$P\colon\mathcal{S}\to\{\mathsf{true},\mathsf{false}\}$ on the set $\mathcal{S}$ of
blockchain states; it must evaluate to $\mathsf{true}$ in the prestate for the
associated execution to succeed.

A settlement transaction and an order each carry a family of precondition predicates.
We write
\[
  \mathcal{P}(\tx)=\{P^1_{\tx},P^2_{\tx},\ldots\}
  \qquad\text{and}\qquad
  \mathcal{P}(\omega)=\{P^1_{\omega},P^2_{\omega},\ldots\}
\]
for the predicates of a transaction $\tx$ and an order $\omega$, respectively. For a
settlement transaction $\tx$, its predicates subsume the order-validation predicates of
every bundled order,
\[
  \mathcal{P}(\tx)\;\supseteq\;\bigcup_{\omega\in\Omega(\tx)}\mathcal{P}(\omega),
\]
because the violation of any such predicate causes $\tx$ to revert. Consequently, $\tx$
succeeds if and only if every predicate holds in its prestate:
\[
  \bigwedge_{P\in\mathcal{P}(\tx)} P\bigl(S_\tx\bigr)=\mathsf{true}.
\]
\end{definition}

\begin{definition}[Failing leg]
\label{def:failing-leg}
The \emph{failing leg} of a reverted settlement transaction $\txr$ is the pair
\[
  \lambda^{\star}(\txr)=(P^{\star}_\txr,\rho^{\star}_\txr),
\]
where $P^{\star}_\txr$ is the first precondition predicate violated along the execution
trace of $\txr$, whose violation causes the revert, and $\rho^{\star}_\txr$ is the
\emph{responsible party} accountable for it. The responsible party 
is defined as the address which can unilaterally flip $P^{\star}_\txr$ from $\mathsf{true}$ to 
$\mathsf{false}$ by submitting an invalidating transaction, as defined below.
\end{definition}

Our threat model (introduced later in Section~\ref{subsec:assumptions}) assumes that prediction market operators 
and other privileged parties are benign. 
In the rest of the paper, we restrict our analysis to failing legs whose responsible 
party $\rho^{\star}_{\txr}$ is a normal prediction market user.
\begin{definition}[Invalidating transaction]
\label{def:invalidating-tx}
An \emph{invalidating transaction} for a reverted settlement transaction $\txr$ is a
committed transaction $\tx<\txr$, submitted by the responsible party
$\rho^{\star}_{\txr}$, that flips the failing predicate $P^{\star}_{\txr}$ from satisfied
to violated:
\[
  P^{\star}_{\txr}(S_{\tx})=\mathsf{true}
  \qquad\text{and}\qquad
  P^{\star}_{\txr}(S^{\prime}_{\tx})=\mathsf{false}.
\]
\end{definition}

Under the same assumption, in the rest of the paper, we focus on 
invalidating transactions 
whose responsible party is a normal user, 
and we call these \emph{order-invalidating transactions}; 
invalidating transactions that could be produced only by an 
operator or another privileged role fall outside our scope.

\begin{definition}[Failing latency]
\label{def:failing-latency}
Let $\tx$ be an order-invalidating transaction for a 
reverted settlement transaction
$\txr$. The \emph{failing latency} of $\txr$ is the blocks between $\tx$ and
$\txr$:
\[
  \Delta(\txr)=\mathsf{Block}(\txr)-\mathsf{Block}(\tx).
\]
\end{definition}

\begin{definition}[Invalidating cost]
\label{def:invalidating-cost}
The \emph{invalidating cost} of a reverted settlement transaction $\txr$ is the gas its
corresponding order-invalidating transaction $\tx$ consumes:
\[
  C(\txr)=\mathsf{GasUsed}(\tx)\cdot\mathsf{GasPrice}(\tx).
\]
\end{definition}

\subsection{Threat Model}
\label{subsec:assumptions}

We consider a non-custodial, blockchain-based prediction market deployed on a
permissionless blockchain. The market exposes two interfaces: an off-chain central
limit order book, and on-chain exchange contracts.

\noindent\textbf{Honest parties.}
We assume that the operator and any other privileged party are \emph{benign}: they
follow the protocol honestly and never intentionally alter order book or settled
transactions, and any deviation on their part is an honest mistake rather than a
deliberate attack. 
We further assume that no privileged party enjoys special access to
the blockchain. 

\noindent\textbf{Adversary.}
Our threat model captures a financially rational adversary $\mathcal{A}$ that is an
ordinary, non-privileged user of the prediction market. 
$\mathcal{A}$ holds at least one
private key for a blockchain account from which it can issue an authenticated
transaction $\tx_{\mathcal{A}}$, and it owns a sufficient balance of the chain's native
cryptocurrency (e.g., \textsc{pol} on Polygon) to perform the actions required by
$\tx_{\mathcal{A}}$.
$\mathcal{A}$ is well connected at the network layer and can observe unconfirmed
transactions in the memory pool. 
It interacts with the market only through the two
exposed surfaces, the off-chain CLOB APIs and the on-chain exchange contracts, and can
neither manipulate the centralized order book nor obtain any privileged access. 
As a non-mining entity, $\mathcal{A}$ influences the relative ordering of its transactions
only by adjusting transaction fees or by resorting to block-building and relay services,
e.g., to front-run~\cite{daian2019flash, torres2021frontrunner} or back-run~\cite{qin2022quantifying} 
a settlement; it cannot unilaterally determine the
contents of a block.

\subsection{Targeted Problems}

We study ghost-filled orders along two complementary directions: (1) 
\emph{analysis},
which reasons about reverted settlement transactions on Polymarket, 
and empirically and quantitatively identifies the responsible party and the economic incentives 
behind them;
and (2) \emph{auditing}, which systematically tests whether existing blockchain-based 
prediction markets are still vulnerable to ghost-filled orders and 
how strongly they incentivize potential attackers. 

\noindent\textbf{Analysis.}
Given a single reverted settlement transaction $\txr$, the analysis problem is to
reconstruct why it failed and who is accountable. Concretely, we seek (i)~its failing
leg $\lambda^{\star}(\txr)=(P^{\star}_{\txr},\rho^{\star}_{\txr})$, the violated
precondition predicate and the responsible party; (ii)~when $\rho^{\star}_{\txr}$ is a
trader, the order-invalidating transaction $\tx$ that flipped $P^{\star}_{\txr}$ from
satisfied to violated, together with the induced failing latency $\Delta(\txr)$; and
(iii)~the economic context of the failure, namely the invalidating cost $C(\txr)$ and the
gain, avoided loss or realized profit, that incentivizes the responsible party.

\noindent\textbf{Auditing.}
Given a blockchain-based prediction market, the auditing problem is to assess its
systemic exposure to ghost-filled orders. This entails
(i)~testing whether it is still possible for a normal user to 
invalidate a settlement transaction by submitting an order-invalidating transaction,
and (ii)~determining whether producing them is economically rational for the
adversary $\mathcal{A}$, i.e., whether the expected gain from invalidating an order
outweighs its cost. The audit thereby characterizes both 
whether the attack can still occur and
how strongly the market incentivizes it.

\section{Analysis Framework and Results}
\label{sec:analysis}

While the existence of ghost-filled orders is known, 
their real-world prevalence, underlying causes, 
and economic incentives remain largely unexplored. 
To study these
questions, we systematically analyze Polymarket's reverted 
settlement transactions. 
For each reverted settlement transaction, 
our goal is to determine:
(i) which settlement precondition was violated and caused the revert,
(ii) which responsible party was able to invalidate that precondition, and
(iii) which earlier transaction actually invalidated it. This attribution enables
us to measure ghost-filled orders at scale and to analyze the incentives of the
parties involved.

\noindent\textbf{Analysis Framework.}
Our analysis framework proceeds in two stages. First, we analyze the smart
contracts used by Polymarket for order settlement
(see Section~\ref{subsec:scope} for details). 
We flatten the relevant contract code and inspect each settlement-related 
function to identify all
revert guards, including \texttt{revert} and
\texttt{require} checks. 
Each guard represents a potential program point at which a
settlement transaction can fail. For every guard, we extract the corresponding
precondition predicate and identify the responsible party, or parties, 
capable of falsifying it, as well as the actions by which they can do so.

Second, we apply this framework to reverted settlement transactions observed in
practice. 
We analyze reverted settlement
transactions in our scope and classify them according to the violated
precondition. 
For preconditions that can be invalidated by a responsible party,
we then backtrack through that responsible party's prior 
transaction history to identify the
specific transaction that flipped the predicate from \texttt{true} to  \texttt{false}. 
The resulting
attribution forms the basis for our subsequent economic analysis of potential
attacker incentives in Section~\ref{subsec:attack-loss-avoidance} and 
Section~\ref{subsec:attack-book-clearing}.

\noindent\textbf{Implementation.}
We realize the analysis framework using two components: \textit{detectors} and
\textit{probes}. Detectors identify the failing leg of a reverted settlement
transaction, while probes recover the earlier transaction that invalidated the
corresponding precondition.
Given a reverted settlement transaction, a detector replays the transaction to
obtain an execution trace. From this trace, it locates the failing leg: the
violated precondition together with the order responsible for triggering that
violation. Once the failing leg is identified, a probe scans the responsible
party's past transaction history to recover the order-invalidating
transaction. We implement one probe for each precondition predicate, using
predicate-specific heuristics to identify the transaction that changed the
predicate from $\mathsf{true}$ to $\mathsf{false}$.

In the next Section~\ref{subsec:preconditions}, 
we use the \texttt{matchOrders} settlement function in the 
Polymarket v1 CTF Exchange contract as a running example 
to illustrate the precondition predicates 
and responsible parties identified from the contract code.
We then describe the setup and scope of our analysis in Section~\ref{subsec:scope} 
and report the coverage of our analysis 
framework in Section~\ref{subsec:coverage}.

\subsection{Precondition Predicates \& Responsible Parties}
\label{subsec:preconditions}

\begin{figure*}[!t]
  \centering
  \includegraphics[width=\textwidth]{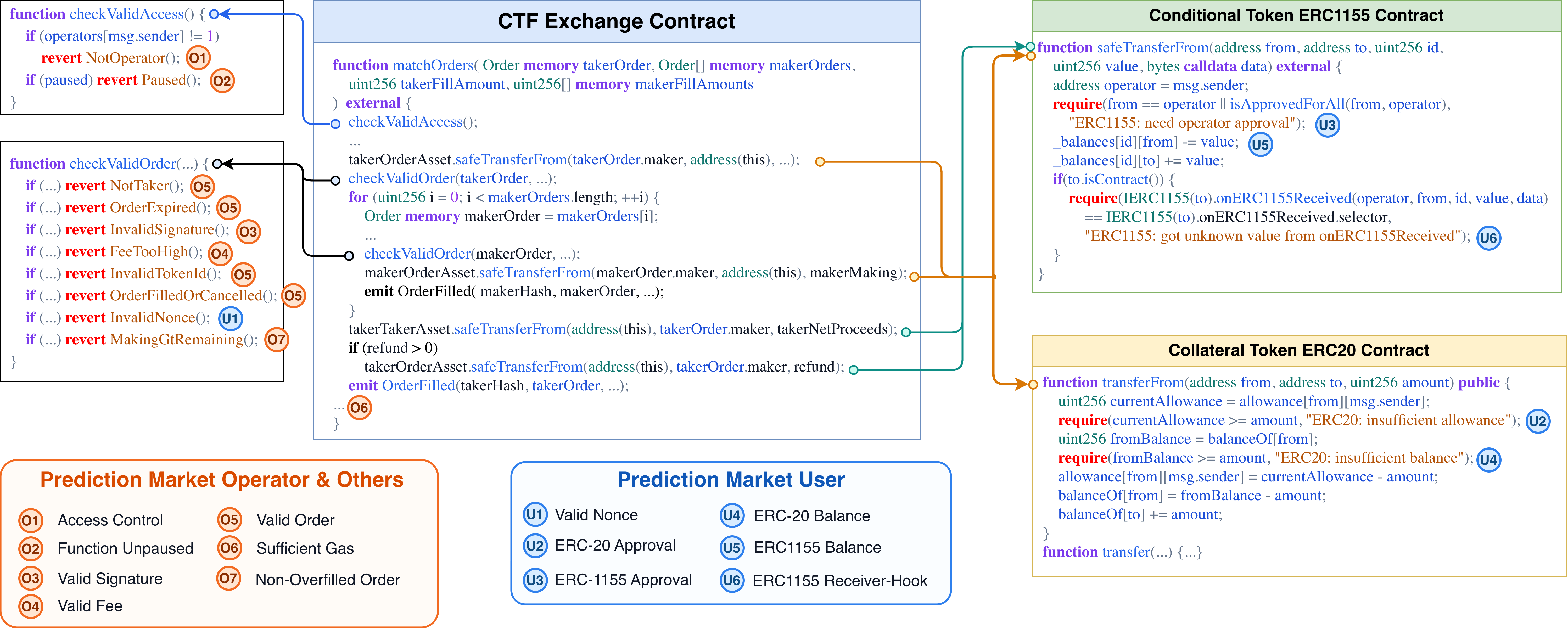}
  \caption{Precondition predicates analyzed for the \texttt{matchOrders} function of the
    Polymarket v1 CTF Exchange contract, together with the responsible party identified
    when each predicate is violated. The code shown is rewritten and simplified for
    presentation; the original code is available at ~\cite{polymarket_ctf_exchange_v1}.}
  \label{fig:preconditions}
\end{figure*}

Figure~\ref{fig:preconditions} illustrates the precondition predicates 
analyzed for the \texttt{matchOrders} function as an example: starting from the
function's source code, we identify every precondition predicate 
whose violation can revert a settlement transaction, 
and we assign to each predicate the party responsible
when it is violated. 

In the following, we enumerate these precondition predicates and, for
each, identify the party responsible when the predicate is violated. 
We organize them by
responsible party: we first present the 
predicates that normal users
cannot invalidate, then those a user can 
control and invalidate (referred to as user-controlled preconditions
in the rest of the paper). 
For each user-controlled precondition, we also explain the 
user actions that can invalidate it, and how the corresponding order-invalidating 
transaction can be detected via category-specific probes.
 Throughout, we use the same notations and definitions mentioned in
Section~\ref{subsec:preliminaries}.

\predcat \textbf{O\nextidx\label{idx:access}: Access Control Precondition.}
The settlement transaction $\txr$ is executable only if its sender is 
an authorized
operator, $P_{\mathsf{access}}(S,\txr) \equiv \mathsf{Authorized}(S,s_{\txr})$, where
$s_{\txr}$ is the sender of $\txr$. 
If $P_{\mathsf{access}}$ is violated, the responsible
party $\rho^{\star}_{\txr}$ is the 
sender $s_{\txr}$, i.e., the operator that submits
$\txr$ or other unauthorized parties that attempt to submit $\txr$. 

\predcat \textbf{O\nextidx\label{idx:active}: Function Unpaused Precondition.}
Neither the contract nor the invoked 
function $f_{\txr}$ may be paused,
$P_{\mathsf{active}}(S,\txr) \equiv \neg\,\mathsf{Paused}(S,f_{\txr})$. The responsible
party $\rho^{\star}_{\txr}$ is the admin that controls the pause state.

\predcat \textbf{O\nextidx\label{idx:signature}: Valid Signature Precondition.}
The transaction-level signature $\sigma_{\txr}$ and every order 
signature in the bundle
$\Omega(\txr)$ must verify, $P_{\mathsf{sig}}(S,\txr) \equiv
\mathsf{VerifyTxSig}(\txr,\sigma_{\txr}) \land \bigwedge_{\omega\in\Omega(\txr)}
\mathsf{VerifyOrderSig}(\omega)$, 
where $\mathsf{VerifyOrderSig}(\omega)$ checks the
signature of $\omega$ against its owner $\mathit{owner}_{\omega}$. 
The responsible party
$\rho^{\star}_{\txr}$ is the operator: it signs $\txr$ and is 
expected to verify every
order signature off-chain before submission, so a single invalid 
order signature reverts
the entire settlement.

\predcat \textbf{O\nextidx\label{idx:fee}: Valid Fee Precondition.}
The fee charged by $\txr$ must not exceed the
exchange contract's maximum fee rate,
$P_{\mathsf{fee}}(S,\txr) \equiv \mathsf{Fee}(\txr) \le \mathsf{MaxFee}(S)$, where
$\mathsf{Fee}(\txr)$ is the fee rate the operator 
specifies in the calldata of $\txr$ and
$\mathsf{MaxFee}(S)$ is the maximum permitted by the 
contract in state $S$. Because the fee
is supplied by the transaction sender in calldata 
rather than by the signed user order,
the responsible party $\rho^{\star}_{\txr}$ is the operator, 
which must choose a fee within
the allowed bound and verify it off-chain before submission 
to avoid overcharging users
and triggering a settlement revert.

\predcat \textbf{O\nextidx\label{idx:valid}: Active Order Precondition.}
For each order $\omega\in\Omega(\txr)$, $\omega$ must be unexpired, 
uncancelled, and not
already filled, formally, $P_{\mathsf{valid}}(S,\txr,\omega) \equiv \mathsf{Time}(\txr) \le
e_{\omega} \land \neg\,\mathsf{Cancelled}(S,\omega) \land \mathsf{Filled}(S,\omega) +
q_{\mathsf{fill}}(\txr,\omega) \le q_{\omega}$,
where $\mathsf{Filled}(S,\omega)$ is the
amount of $\omega$ already filled and $q_{\mathsf{fill}}(\txr,\omega)$ 
the amount $\txr$
attempts to fill. 
The responsible party $\rho^{\star}_{\txr}$ is the operator, which should 
verify expiry, cancellation, and the remaining fillable amount off-chain before
submission.

\predcat \textbf{O\nextidx\label{idx:gas}: Sufficient Gas Precondition.}
The transaction must supply enough gas to 
execute within the block gas limit, formally, 
$P_{\mathsf{gas}}(\txr) \equiv \mathit{gas}_{\txr} \ge \mathsf{GasNeeded}(\txr)$,
 where
$\mathit{gas}_{\txr}$ is the gas limit of $\txr$ and $\mathsf{GasNeeded}(\txr)$ the gas
its execution requires, 
which grows with the number of orders in $\Omega(\txr)$. The
responsible party $\rho^{\star}_{\txr}$ is the operator, which should 
have supplied enough gas for the transaction to execute, and should control 
the number of orders included in the bundle to avoid exceeding the block gas limit.\footnote{If an ERC-1155
receiver hook is written to consume infinite gas, e.g., through an infinite loop,
the gas requirement is controlled by the order's recipient rather than the operator; in
that case the responsible party is the user who controls the hook code and
configuration.}

\predcat \textbf{O\nextidx\label{idx:nooverfill}: Non-Overfilled Order Precondition.}
Orders in $\Omega(\txr)$ that draw on the same user 
collateral or conditional token must
not jointly demand more than is available: 
for each token $t$ (an ERC-20 balance or
allowance, or an ERC-1155 balance) and the orders $\Omega_t(\txr)\subseteq\Omega(\txr)$
that consume it, $P_{\mathsf{nooverfill}}(S,\txr) \equiv \bigwedge_{t}
\sum_{\omega\in\Omega_t(\txr)} q_{\mathsf{fill}}(\txr,\omega) \le \mathsf{Avail}_t(S)$,
where $\mathsf{Avail}_t(S)$ is the amount of $t$ available in state $S$. In this case, 
each order is completely
valid individually at the prestate, 
while the bundle is not, 
when the cumulative demand exceeds the
available balance or allowance.  
It typically happens when a user submits multiple
orders in the same direction, 
and the operator fills them together in a single settlement transaction, 
while actually only a subset of them can be filled.
The responsible party $\rho^{\star}_{\txr}$ is the
operator, which controls both the orders included 
in $\Omega(\txr)$ and should have checked
the cumulative fill against the available balance or allowance for 
each token to ensure that the bundle is executable, 
and can therefore avoid overfilling any token.

\predcat \textbf{Precondition Identification for O1-O7:}
Given a reverted settlement transaction $\txr$, 
we identify the violated precondition as
follows. 
We replay $\txr$, capture the revert data from its execution trace, 
and decode it into a revert selector. 
When this selector matches a corresponding known revert reason, we attribute
the failure directly to O\ref{idx:access}, O\ref{idx:active}, 
O\ref{idx:signature}, O\ref{idx:fee} or
O\ref{idx:valid}. 
For O\ref{idx:gas}, we check whether the gas consumed is very close to the
gas limit; if so, and provided the gas exhaustion is 
not caused by an ERC-1155 receiver hook,
we attribute the revert to O\ref{idx:gas}. 
For O\ref{idx:nooverfill}, we first decode the calldata 
to recover the orders in $\Omega(\txr)$, and then we 
confirm
that every order in the bundle is individually 
valid and backed by sufficient balance and
approval in the prestate $S_{\txr}$; 
we then compare, for each token, the cumulative fill
of the orders that consume it against the 
available balance or allowance in $S_{\txr}$,
and attribute the revert to O\ref{idx:nooverfill} 
if any token is overfilled.

\predcat \textbf{U\nextidxU\label{index:nonce}: Valid Nonce Precondition.}
In Polymarket v1, each signed order has a nonce that is 
checked at settlement against
the order owner's current on-chain account nonce, 
so the order is executable only if the
two agree, $P_{\mathsf{nonce}}(S,\omega) \equiv n_{\omega} =
\mathsf{Nonce}(S,\mathit{owner}_{\omega})$, 
where $n_{\omega}$ is the order's nonce and
$\mathsf{Nonce}(S,a)$ the on-chain account nonce of 
account $a$ in state $S$. The nonce
written into a signed order is immutable. 
The owner can, however, freely increment
$\mathsf{Nonce}(S,\mathit{owner}_{\omega})$ by invoking
a function ``incrementNonce'' on the Polymarket account contracts,
and doing 
so invalidates every order they
previously signed under the old nonce, including 
both maker and taker orders.\footnote{This
mechanism was removed in Polymarket v2, so it 
applies only to the v1 contracts, which are
no longer active. As of this writing (June 9, 2026), 
a v1 account is automatically banned
if it ever increments its nonce.} 
The responsible party $\rho^{\star}_{\txr}$ is therefore
the order owner, who can invalidate the order at 
will by incrementing their account nonce
before the settlement transaction executes.

\emph{Backtracking the order-invalidating 
transaction for U\ref{index:nonce}.}
To recover this transaction, we 
replay the reverted settlement 
transaction $\txr$ to
identify the order owner $\mathit{owner}_{\omega}$ 
and the nonce its order was signed
under, and then search that account's history 
for the most recent \texttt{incrementNonce}
call preceding $\txr$. That call is considered the order-
invalidating transaction: it increments the 
account nonce, so the
order no longer satisfies $P_{\mathsf{nonce}}$.

\predcat \textbf{U\nextidxU\label{index:erc20_approval}: 
ERC-20 Approval Precondition} and
\textbf{U\nextidxU\label{index:erc1155_approval}: 
ERC-1155 Approval Precondition.}
As Figure~\ref{fig:preconditions} shows, 
every \texttt{transferFrom} of an ERC-20 or
ERC-1155 token in a settlement 
transaction $\txr$ is guarded by an approval check, so
$\txr$ is executable only if the operator is authorized 
to move all of the tokens needed
to settle the orders in $\Omega(\txr)$. 
Consider a transfer that moves $q$ units of token
$\gamma$ from an order owner $a$ on the operator $o$'s behalf. 
For ERC-20 collateral, the
allowance must cover the transfer, $P^{20}_{\mathsf{allow}}(S) \equiv
\mathsf{Allow}^{20}_{\gamma}(S,a,o) \ge q$; 
for ERC-1155 conditional tokens, the operator
must hold blanket approval, $P^{1155}_{\mathsf{appr}}(S) \equiv
\mathsf{ApprovedForAll}^{1155}_{\gamma}(S,a,o)$, 
granted through \texttt{setApprovalForAll}.
The responsible party $\rho^{\star}_{\txr}$ is the order 
owner, who can invalidate the
order by lowering $\mathsf{Allow}^{20}_{\gamma}(S,a,o)$ below $q$, 
or by revoking the
operator's ERC-1155 approval, at any point 
before the settlement transaction.

\predcat \textbf{U\nextidxU\label{index:erc20_balance}: ERC-20 Balance Precondition} and
\textbf{U\nextidxU\label{index:erc1155_balance}: ERC-1155 Balance Precondition.}
Similarly, beyond approval, every \texttt{transferFrom} also 
verifies that the debited account
actually holds the tokens being moved, also as shown 
in Figure~\ref{fig:preconditions}. 
For the
same transfer of $q$ units of token $\gamma$ from account $a$
(bearing token id $\iota$ in
the ERC-1155 case), 
the preconditions are $P^{20}_{\mathsf{bal}}(S) \equiv
\mathsf{Bal}^{20}_{\gamma}(S,a) \ge q$ for ERC-20 
collateral and $P^{1155}_{\mathsf{bal}}(S)
\equiv \mathsf{Bal}^{1155}_{\gamma}(S,a,\iota) \ge q$ for 
ERC-1155 conditional tokens, 
both evaluated in the prestate $S_{\txr}$. 
Hence $\txr$ is executable only if every account
holds enough tokens to cover the transfers 
required to settle the orders in $\Omega(\txr)$.
The responsible party $\rho^{\star}_{\txr}$ is 
again the order owner, who can invalidate the
order by transferring the relevant 
tokens out of the account, driving
$\mathsf{Bal}^{20}_{\gamma}$ or 
$\mathsf{Bal}^{1155}_{\gamma}$ below $q$ before the
settlement transaction executes.

\predcat \textbf{U\nextidxU\label{index:erc1155_hook}: 
ERC-1155 Receiver Hook Precondition.}
As Figure~\ref{fig:preconditions} shows, an 
ERC-1155 \texttt{transferFrom} whose recipient
$r$ is a contract automatically 
invokes the recipient's \texttt{onERC1155Received} hook,
and the transfer succeeds only if the hook 
returns the expected acceptance value, formally,
$P^{1155}_{\mathsf{hook}}(S) \equiv \neg \text{isContract}(r) \lor
\mathsf{onERC1155Received}(S,r) = \mathsf{ACCEPT}$. 
Because some Polymarket account types are
themselves contracts, their owners 
can configure the hook's code and parameters to accept
or reject incoming transfers.\footnote{As of this writing 
(June 9, 2026), Polymarket also
automatically bans accounts that update their 
account contract to add or change hook code,
so this attack vector is only theoretical: 
it requires updating the hook code and reverting 
the settlement transaction before the account is banned.} 
The responsible party
$\rho^{\star}_{\txr}$ is the order owner, who can 
invalidate the order by reprogramming the
\texttt{onERC1155Received} hook so 
that $P^{1155}_{\mathsf{hook}}$ fails. It can 
either rejects the incoming transfer or consumes unbounded gas 
in an infinite loop, so that the settlement
transaction reverts when it attempts 
to deliver ERC-1155 tokens to that account.

\predcat \textbf{Failing Leg Identification for U1-U6:}
Unlike O1-O7, U1-U6 requires identifying which order's 
precondition is violated and attribute the failure to a 
specific responsible party
to support later analysis in 
Section~\ref{subsec:attack-loss-avoidance} 
and Section~\ref{subsec:attack-book-clearing}.
For a reverted settlement transaction $\txr$ not attributed to O1-O7, 
we replay $\txr$ to obtain its execution trace.
We decode the calldata using the function signature to 
extract the order list, then analyze \texttt{OrderFilled} events from the execution trace.
By examining the last \texttt{OrderFilled} event and the 
revert data, we determine the current execution step.
For example, in Polymarket v1 CTF Exchange (\texttt{matchOrders}), 
maker orders are filled sequentially before the taker order (Figure~\ref{fig:preconditions}).
If the revert data matches U\ref{index:nonce}'s revert reason, 
the violating order has an invalid nonce.
If it matches U\ref{index:erc20_approval}-U\ref{index:erc1155_balance}, 
we identify the violated precondition by examining which 
order is being filled, its side (BUY/SELL), and the token involved.
Otherwise, we need to 
trace the call stack of the execution trace to
detect ERC1155 receiver hook invocations. 
If the revert occurs 
within the hook,  we attribute the revert to U\ref{index:erc1155_hook}, 
and the responsible party is the order 
owner who controls the hook. After failing leg identification,
our probes can then backtrack order-invalidating transactions as described below.

\predcat \textbf{Order-invalidating Transaction Backtracking for
U\ref{index:erc20_approval}-U\ref{index:erc1155_balance}:}
For these four predicates, we first replay 
the reverted settlement transaction $\txr$ to
determine which balance or approval 
check failed and by how much, that is, the token
$\gamma$, the account $a$, and the required amount $q$. 
We then search that account's
latest transaction 
that flipped the corresponding predicate from
\textit{true} to \textit{false} before $\txr$.
For balance failures ($P^{20}_{\mathsf{bal}}$,
$P^{1155}_{\mathsf{bal}}$), we reconstruct the 
relevant balance from the token contract's
\texttt{Transfer}, \texttt{TransferSingle}, 
and \texttt{TransferBatch} events and locate
the transaction at which $\mathsf{Bal}^{20}_{\gamma}(S,a)$ or
$\mathsf{Bal}^{1155}_{\gamma}(S,a,\iota)$ 
first dropped below $q$. 
For ERC-20 approval
failures ($P^{20}_{\mathsf{allow}}$), we 
scan \texttt{Approval} events together with the
allowance-consuming \texttt{transferFrom} calls 
to find where
$\mathsf{Allowance}^{20}_{\gamma}(S,a,o)$ fell below $q$. 
For ERC-1155 approval failures
($P^{1155}_{\mathsf{appr}}$), we scan \texttt{ApprovalForAll} 
events for the most recent
revocation that set $\mathsf{ApprovedForAll}^{1155}_{\gamma}(S,a,o)$ 
to $\mathsf{false}$.
In each case, the latest such transaction 
preceding $\txr$ is the order-invalidating
transaction.\footnote{We implement these probes using 
Google BigQuery and Polygon RPC methods. 
Because scanning an account's full 
transaction history can be prohibitively 
expensive, if the probe does not 
find order-invalidating transactions within a reasonable 
number of recent blocks,
we instead binary-search 
over blocks with Polygon RPC calls to 
find the block at which the allowance or balance flipped, 
and then examine the transactions in that block to pinpoint 
the exact order-invalidating transaction.}

\predcat \textbf{Order-invalidating Transaction Backtracking for
U\ref{index:erc1155_hook}:}
We first replay the reverted settlement 
transaction $\txr$ to confirm that it failed on
an ERC-1155 \texttt{safeTransferFrom} or 
\texttt{safeBatchTransferFrom} leg whose token
transfer reached the recipient $r$ but whose 
receiver-side acceptance check failed,
typically because \texttt{onERC1155Received} or 
\texttt{onERC1155BatchReceived} reverted,
returned an invalid selector, or exhausted 
gas in an infinite loop. We then search the
recipient's history for the prior state change 
that rendered it unable to accept ERC-1155
transfers, i.e., the transaction that 
changed $P^{1155}_{\mathsf{hook}}$ to
$\mathsf{false}$, such as the creation or upgrade 
of the receiver contract, an EIP-7702
delegation change, etc. The latest such transaction
preceding $\txr$ is considered the order-invalidating transaction.

\subsection{Analysis Setup and Scope}
\label{subsec:scope}

We analyze ghost-filled orders by analyzing 
reverted settlement transactions on
Polymarket. We collect every reverted 
transaction that invokes one of the 9 settlement
functions across the 5 contracts 
listed in Table~\ref{tab:studied-contracts}; these are
the only functions in the Polymarket contracts 
through which an operator can settle an
order on-chain. 
Our dataset spans 2025-08-12 to 2026-05-22 (UTC) and comprises
1,824,926 reverted transactions in total.

\begin{table}[!ht]
  \centering
  \caption{Polymarket contracts and functions analyzed.}
  \label{tab:studied-contracts}
  \scriptsize
  \setlength{\tabcolsep}{4pt}
  \begin{tabular}{@{}ll@{}}
    \toprule
    \textbf{Contract} & \textbf{Functions analyzed} \\
    \midrule
    \texttt{CTF\_EXCHANGE\_V1}~\cite{polymarket_ctf_exchange_v1}            & \texttt{matchOrders}, \texttt{fillOrder(s)}$^\dagger$ \\
    \texttt{NEGRISK\_CTF\_EXCHANGE\_V1}~\cite{polymarket_negrisk_ctf_exchange_v1} & \texttt{matchOrders}, \texttt{fillOrder(s)}$^\dagger$ \\
    \texttt{V1\_FEE\_MODULE}~\cite{polymarket_fee_module_v1}              & \texttt{matchOrders}$^\ddagger$ \\
    \texttt{CTF\_EXCHANGE\_V2}~\cite{polymarket_ctf_exchange_v2}            & \texttt{matchOrders} \\
    \texttt{NEGRISK\_CTF\_EXCHANGE\_V2}~\cite{polymarket_negrisk_ctf_exchange_v2} & \texttt{matchOrders} \\
    \bottomrule
  \end{tabular}
  \par\smallskip
  \begin{minipage}{\linewidth}
    \raggedright\footnotesize
    $^\dagger$\texttt{fillOrder(s)} denotes both \texttt{fillOrder} and
    \texttt{fillOrders}. $^\ddagger$Fee-module variant only.
  \end{minipage}
\end{table}

\subsection{Analysis Coverage, Invalidating Cost \& Latency}
\label{subsec:coverage}

\newcommand{\LatencyHistNonce}{%
  \begin{tikzpicture}[x=3.0pt,y=2.0pt,baseline=-0.85ex]
    \draw[black,line width=0.2pt] (-0.35,0) -- (23.35,0);
    \foreach \x/\h in {0/0.945, 1/0.856, 2/0.300, 3/0.151, 4/0.150, 5/0.150, 6/0.150, 7/0.150, 8/0.322, 9/2.934, 10/7.500, 11/0.000} {%
      \fill[black] ({2*\x},0) rectangle ++(1.20,\h);
    }
    \foreach \x/\lab in {0.60/0,2.60/1,4.60/2,7.20/5,9.20/10,11.20/20,13.20/50,15.20/100} {%
      \draw[black,line width=0.15pt] (\x,0) -- (\x,-0.35);
      \node[font=\tiny,anchor=north east,rotate=35,inner sep=0.2pt] at (\x,-0.48) {\lab};
    }
  \end{tikzpicture}%
}

\newcommand{\LatencyHistERCtwentyApproval}{%
  \begin{tikzpicture}[x=3.0pt,y=2.0pt,baseline=-0.85ex]
    \draw[black,line width=0.2pt] (-0.35,0) -- (23.35,0);
    \foreach \x/\h in {0/3.560, 1/0.528, 2/0.150, 3/0.539, 4/0.202, 5/0.150, 6/0.150, 7/0.150, 8/0.157, 9/1.011, 10/5.183, 11/1.715} {%
      \fill[black] ({2*\x},0) rectangle ++(1.20,\h);
    }
    \foreach \x/\lab in {0.60/0,2.60/1,4.60/2,7.20/5,9.20/10,11.20/20,13.20/50,15.20/100} {%
      \draw[black,line width=0.15pt] (\x,0) -- (\x,-0.35);
      \node[font=\tiny,anchor=north east,rotate=35,inner sep=0.2pt] at (\x,-0.48) {\lab};
    }
  \end{tikzpicture}%
}

\newcommand{\LatencyHistERCelevenfiftyfiveApproval}{%
  \begin{tikzpicture}[x=3.0pt,y=2.0pt,baseline=-0.85ex]
    \draw[black,line width=0.2pt] (-0.35,0) -- (23.35,0);
    \foreach \x/\h in {0/4.707, 1/1.367, 2/1.367, 3/0.000, 4/0.304, 5/0.152, 6/1.367, 7/0.607, 8/1.974, 9/1.215, 10/0.000, 11/0.000} {%
      \fill[black] ({2*\x},0) rectangle ++(1.20,\h);
    }
    \foreach \x/\lab in {0.60/0,2.60/1,4.60/2,7.20/5,9.20/10,11.20/20,13.20/50,15.20/100} {%
      \draw[black,line width=0.15pt] (\x,0) -- (\x,-0.35);
      \node[font=\tiny,anchor=north east,rotate=35,inner sep=0.2pt] at (\x,-0.48) {\lab};
    }
  \end{tikzpicture}%
}

\newcommand{\LatencyHistERCtwentyBalance}{%
  \begin{tikzpicture}[x=3.0pt,y=2.0pt,baseline=-0.85ex]
    \draw[black,line width=0.2pt] (-0.35,0) -- (23.35,0);
    \foreach \x/\h in {0/6.034, 1/3.334, 2/1.278, 3/0.397, 4/0.173, 5/0.150, 6/0.172, 7/0.180, 8/0.706, 9/0.502, 10/0.157, 11/0.150} {%
      \fill[black] ({2*\x},0) rectangle ++(1.20,\h);
    }
    \foreach \x/\lab in {0.60/0,2.60/1,4.60/2,7.20/5,9.20/10,11.20/20,13.20/50,15.20/100} {%
      \draw[black,line width=0.15pt] (\x,0) -- (\x,-0.35);
      \node[font=\tiny,anchor=north east,rotate=35,inner sep=0.2pt] at (\x,-0.48) {\lab};
    }
  \end{tikzpicture}%
}

\newcommand{\LatencyHistERCelevenfiftyfiveBalance}{%
  \begin{tikzpicture}[x=3.0pt,y=2.0pt,baseline=-0.85ex]
    \draw[black,line width=0.2pt] (-0.35,0) -- (23.35,0);
    \foreach \x/\h in {0/4.075, 1/4.539, 2/2.514, 3/0.599, 4/0.309, 5/0.183, 6/0.183, 7/0.150, 8/0.290, 9/0.200, 10/0.150, 11/0.150} {%
      \fill[black] ({2*\x},0) rectangle ++(1.20,\h);
    }
    \foreach \x/\lab in {0.60/0,2.60/1,4.60/2,7.20/5,9.20/10,11.20/20,13.20/50,15.20/100} {%
      \draw[black,line width=0.15pt] (\x,0) -- (\x,-0.35);
      \node[font=\tiny,anchor=north east,rotate=35,inner sep=0.2pt] at (\x,-0.48) {\lab};
    }
  \end{tikzpicture}%
}

\newcommand{\LatencyHistERCelevenfiftyfiveHook}{%
  \begin{tikzpicture}[x=3.0pt,y=2.0pt,baseline=-0.85ex]
    \draw[black,line width=0.2pt] (-0.35,0) -- (23.35,0);
    \foreach \x/\h in {0/0.260, 1/0.665, 2/1.053, 3/2.323, 4/4.601, 5/3.771, 6/0.299, 7/0.150, 8/0.150, 9/0.150, 10/0.150, 11/0.150} {%
      \fill[black] ({2*\x},0) rectangle ++(1.20,\h);
    }
    \foreach \x/\lab in {0.60/0,2.60/1,4.60/2,7.20/5,9.20/10,11.20/20,13.20/50,15.20/100} {%
      \draw[black,line width=0.15pt] (\x,0) -- (\x,-0.35);
      \node[font=\tiny,anchor=north east,rotate=35,inner sep=0.2pt] at (\x,-0.48) {\lab};
    }
  \end{tikzpicture}%
}

\begin{table*}[!t]
\centering
\caption{Per-precondition invalidating cost and failing latency, with the latency distribution.}
\label{tab:stats}
\footnotesize
\setlength{\tabcolsep}{4pt}
\begin{tabular}{lrrrrr}
\toprule
Precondition & \shortstack[r]{Reverted\\Transactions} & \shortstack[r]{Order-invalidating\\Transactions Found} & \shortstack[r]{Avg. Failing\\Cost (USD)} & \shortstack[r]{Avg. Failing\\Latency (blocks)} & \shortstack[r]{Failing Latency\\Histogram} \\
\midrule
O1--O7: Operator-side & 50,271 & -- & -- & -- & -- \\
U\ref{index:nonce}: Order Valid Nonce Precondition & 19,882 & 19,882 & 0.0062 & 15,465.4 & \LatencyHistNonce \\
U\ref{index:erc20_approval}: ERC-20 Approval Precondition & 63,884 & 63,882 & 0.0183 & 37,409.7 & \LatencyHistERCtwentyApproval \\
U\ref{index:erc1155_approval}: ERC-1155 Approval Precondition & 95 & 86 & 0.0010 & 229.9 & \LatencyHistERCelevenfiftyfiveApproval \\
U\ref{index:erc20_balance}: ERC-20 Balance Precondition & 662,867 & 649,589 & 0.0193 & 694.7 & \LatencyHistERCtwentyBalance \\
U\ref{index:erc1155_balance}: ERC-1155 Balance Precondition & 255,832 & 208,806 & 0.0213 & 148.6 & \LatencyHistERCelevenfiftyfiveBalance \\
U\ref{index:erc1155_hook}: ERC-1155 Receiver Hook Precondition & 772,095 & 771,832 & 0.0149 & 1,562.7 & \LatencyHistERCelevenfiftyfiveHook \\
\midrule
Total & 1,824,926 & 1,714,077 & 0.0174 & 2,558.7 & -- \\
\bottomrule
\end{tabular}

\end{table*}

Table~\ref{tab:stats} reports the reverted settlement 
transactions in our scope, grouped by the precondition 
violated by the reverted transaction.
We distinguish two notions of coverage. 
The first is \emph{classification coverage}: 
whether every reverted settlement transaction in
our dataset can be assigned to one of the preconditions in 
our taxonomy in Section~\ref{subsec:preconditions}. 
The second is \emph{attribution coverage}: 
for reverted transactions violating 
user-controlled preconditions, 
whether our probes can identify a 
concrete order-invalidating transaction 
that caused the later settlement transaction to fail. 
The operator-side preconditions O1--O7 
are not user-controlled, and therefore we do not 
attempt to backtrack them. 
For each pair of reverted transaction and 
order-invalidating transaction, 
we also 
compute the \emph{failing latency}, 
defined as the block distance between 
them. 
A latency of zero means that the invalidating 
transaction is included ahead of 
the failed settlement transaction, but in the same block, 
probably due to front-running by the responsible party.
We compute the \emph{invalidating cost} as the gas 
consumed by the order-invalidating transaction 
multiplied by its effective gas price and the 
USD price of the chain-native gas token 
(i.e., POL on Polygon).

\noindent\textbf{Analysis Coverage.}
Table~\ref{tab:stats} presents the results of our analysis. 
Our precondition taxonomy in Section~\ref{subsec:preconditions} successfully 
covers all reverted settlement transactions in the scope. 
The overwhelming majority of preconditions violated are user-controlled. 
The operator-side preconditions O1--O7 account for only 
50,271 transactions, or 2.8\% of all reverts, 
whereas U1--U6 account for 1,774,655 transactions, or 97.2\%. 
This suggests that reverted settlement transactions 
are primarily caused by preconditions that 
depend on user-controlled on-chain state, 
such as token balances.
Out of these 1,774,655 transactions, 
our probes successfully identify 
1,714,077 order-invalidating transactions, 
giving an overall attribution coverage of 96.6\%. 
Coverage is complete for the nonce (U\ref{index:nonce}), 
near-complete for
ERC-20 approval (U\ref{index:erc20_approval}), 
and ERC-1155 hook (U\ref{index:erc1155_hook}) preconditions, 
and high for ERC-20 balance (U\ref{index:erc20_balance}). 
 The main source of missed attribution is the ERC-1155 
 balance precondition U\ref{index:erc1155_balance}: 
 our probes left 47,026 unattributed cases. 
 This row alone accounts for 77.6\% of all 
 unattributed user-controlled reverts.
 These misses result from a bounded backward search. 
 ERC-1155 tokens are harder to track than ERC-20
 tokens, because they can be merged or split. 
 An exhaustive search for ERC-1155 balance changes 
 across the entire transaction history of the relevant 
 accounts would be very expensive, and our probes
 only search within a fixed window of recent blocks 
 before the reverted transaction.

\noindent\textbf{Invalidating Cost.}
The average invalidating cost is very small across all preconditions.\footnote{
  In practice, people also need to prepare initial 
  balances and approvals before conducting ghost-filled 
  orders, but these costs are very hard to track. 
  In fact, since these operations are not 
  time-sensitive, they can wait until the gas price
  is low to execute, so the cost can be very low. 
} 
The overall average is only 0.0174 USD per reverted settlement transaction. 
The cheapest category is the ERC-1155 approval precondition
 U\ref{index:erc1155_approval}, at 0.0010 USD, but only 
 95 transactions belong to this category. 
 The second cheapest is the nonce precondition 
 U\ref{index:nonce}, at 0.0062 USD, which could also 
 explain why Polymarket emphasized it in their public announcement,
 and fixed it soon in v2.

\noindent\textbf{Failing Latency.}
The average failing latency is quite long, at 2,558.7 blocks, 
or about 1.2 hours on Polygon.
The latency histograms provide a stronger signal 
than the averages, because the averages are skewed 
by long-tail stale-state cases, which supports the finding that 
not all of the ghost-filled orders are the result of strategic 
invalidation, but some of them are simply stale 
orders, operational mistakes, or just an implementation bug. 
A long latency may correspond to an old cancellation, 
an old approval revocation, or an account 
that became undercollateralized long before settlement. 
By contrast, a zero-block or one-block latency 
indicates that the invalidation is tightly coupled to 
the failed settlement transaction, which is the pattern 
expected from strategic order invalidation. 
This pattern is especially visible for U2--U5:
balance and approval preconditions. For example,
for ERC-20 balance failures, 46.2\% of located invalidations 
occur in the same block as the reverted settlement transaction, 
and 71.7\% occur within one block. 
These short-latency clusters are difficult to explain 
solely as accidental stale orders, and are consistent 
with users or automated agents invalidating 
settlement transactions immediately before or 
near the settlement attempt.

\subsection{Case Study 1: Avoiding Losses After Market Closure}
\label{subsec:attack-loss-avoidance}

While ghost-filled orders are a known issue~\cite{itslirrato_2026} 
caused by the lack of 
atomicity in
non-custodial prediction markets, it remains unclear 
whether they have been
widely exploited for malicious purposes or 
unjustified trader benefits. 
We identify two highly suspicious 
scenarios in which ghost-filled orders 
can benefit traders unfairly. 
To measure their prevalence and 
financial incentives, we
implemented analysis tools and applied 
them to all markets within the scope
specified in Section~\ref{subsec:scope}.

As discussed in Section~\ref{sec:ghost-filled-orders}, ghost-filled orders give
the responsible party an unusually blessed option: 
they can cancel an order anytime before
it settles on-chain. 
This allows the owner to observe how the market moves and
then decide whether settlement should succeed. 
If the order would make money,
the owner can let it settle; otherwise, the owner can invalidate one of the
settlement preconditions and cause the transaction to revert. In other words,
the owner gets a ``free look'' at the market outcome: 
heads, they settle; tails, they revert.

We illustrate this strategy using an example of 
BTC Up/Down 5-minute market, where the
outcome depends on whether the bitcoin price goes up or down by the end of the
next 5-minute interval. 
Suppose the market resolves to \textsc{Up}. An order
that would buy \textsc{Down} would lose money if settled, 
so reverting it
\emph{avoids a loss}. In contrast, an order that would buy \textsc{Up} would
make money if settled, so reverting it \emph{forgoes a gain}. 
If reverts were
mostly accidental, avoided losses and forgone gains should both appear; the
distribution should not overwhelmingly favor one side.

\begin{figure}[t]
  \centering
  \includegraphics[width=0.7 \columnwidth]{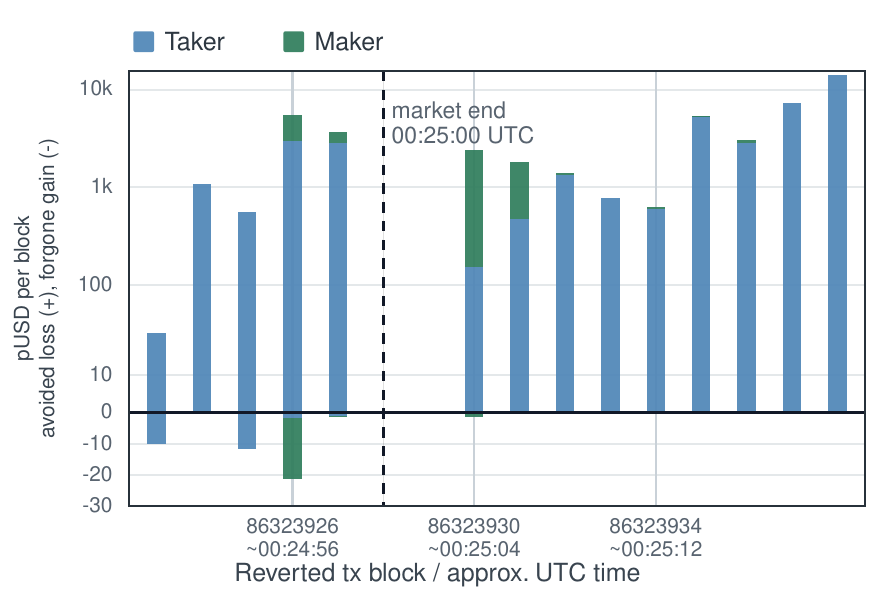}
  \caption{Post-mortem of the Bitcoin Up and Down 5m market (UTC00:20--00:25) on May 2, 2026: avoided
    loss and forgone gain across its reverted transactions.}
  \label{fig:btc-postmortem}
\end{figure}

Figure~\ref{fig:btc-postmortem} shows the economic 
outcome of reverted
transactions near and after the end of the BTC Up/Down 5-minute market from
UTC00:20 to UTC00:25 on May 2, 2026. 
For each failing leg, we perform a
post-mortem using the final market outcome.  
We define the \emph{avoided loss} as
the amount the order owner would have lost had the 
order settled, and the
\emph{forgone gain} as the amount the order owner would 
have gained. 
The result is surprisingly strongly one-sided: 
most reverted transactions avoided losses. 
Before the market ended, a small fraction still forgone gains, likely due to timing mistakes
or uncertainty about the final outcome. 
After the market ended, however, almost
all reverted transactions avoided losses. 
It can be explained by 
the fact that after the market ends, the outcome is 
already known to the public, 
so the strategy becomes almost risk-free: 
revert losing orders and keep winning ones.

\begin{table}[t]
\centering
\caption{Financial impact by failing-leg side 
and reverted transaction timing for reverted 
transactions whose identified order-invalidating transactions
occur within 60 seconds before market end or
anytime after market end.}
\label{tab:side-revert-boundary-financial-impact}
\footnotesize
\setlength{\tabcolsep}{3pt}
\begin{tabular}{@{}llrrrrr@{}}
\toprule
Role & Timing
  & \shortstack[r]{\# Txs}
  & \shortstack[r]{Avoided\\Loss}
  & \shortstack[r]{Forgone\\Gain}
  & \shortstack[r]{Loss/Gain\\ratio}
  & \shortstack[r]{Avoided\\Loss rate} \\
\midrule
Maker & $\le$ end & 133,026 & 2,789,524  & 207,213 & 13:1   & 85\% \\
Taker & $\le$ end & 89,688  & 9,081,892  & 121,066 & 75:1   & 81\% \\
Maker & $>$ end   & 299,604 & 10,853,501 & 42,163  & 257:1  & 99\% \\
Taker & $>$ end   & 235,262 & 38,554,570 & 9,980   & 3863:1 & 98\% \\
\midrule
Total & -- & 757,580 & 61,279,487 & 380,422 & 161:1 & 94\% \\
\bottomrule
\end{tabular}
\end{table}

Table~\ref{tab:side-revert-boundary-financial-impact}
further extends the analysis to reverted transactions
whose identified order-invalidating transactions
occur within 60 seconds before market end or
anytime after market end, within our scope, for 
a total of 757,580 transactions.
The \emph{loss/gain ratio} is the ratio
between avoided loss and forgone gain, 
and the \emph{avoided-loss rate} is the
fraction of reverted transactions that avoided a loss.

Across  757,580 reverted transactions, 94\% avoided a loss, 
and the total avoided loss exceeds the total forgone 
gain by 161:1. This imbalance is extremely 
difficult to
explain as random failure. 
Timing further strengthens the evidence. Before the
market ends, the avoided-loss rate is high but 
imperfect: 85\% for makers and
81\% for takers. After the market ends, when the 
outcome is effectively known,
the avoided-loss rate rises to 99\% for makers and 
98\% for takers, making the
revert a highly reliable escape hatch.
The effect is also stronger for taker orders 
than for maker orders. We attribute
this to settlement timing. 
A maker order rests in the book, and the moment it is
matched and finalized is hard to anticipate. 
A taker order, by contrast,
immediately consumes a resting maker order, 
typically through a fill-or-kill
order, so its owner can better predict when 
settlement will occur and time the
revert accordingly.

Finally, avoided losses are highly concentrated in 
short-horizon crypto up/down (5-minute, 15-minute, and 4-hour)
markets. 
Four types of crypto up/down markets account
for 78\% of the total avoided loss: BTC (\$17.0M), ETH (\$11.3M),
SOL (\$9.9M), and XRP (\$9.6M). 
This concentration suggests that
reverting settlement transactions has 
become a deliberate strategy tailored to
volatile, high-frequency  markets such as crypto up/down, 
where last-second price movements make profitable reverts far more common.

\subsection{Case Study 2: Order-Book Clearing and 
Maker Arbitrage Opportunities
in the Middle of an Active Market}
\label{subsec:attack-book-clearing}

\begin{figure}[t]
    \centering
    \includegraphics[width=\columnwidth]{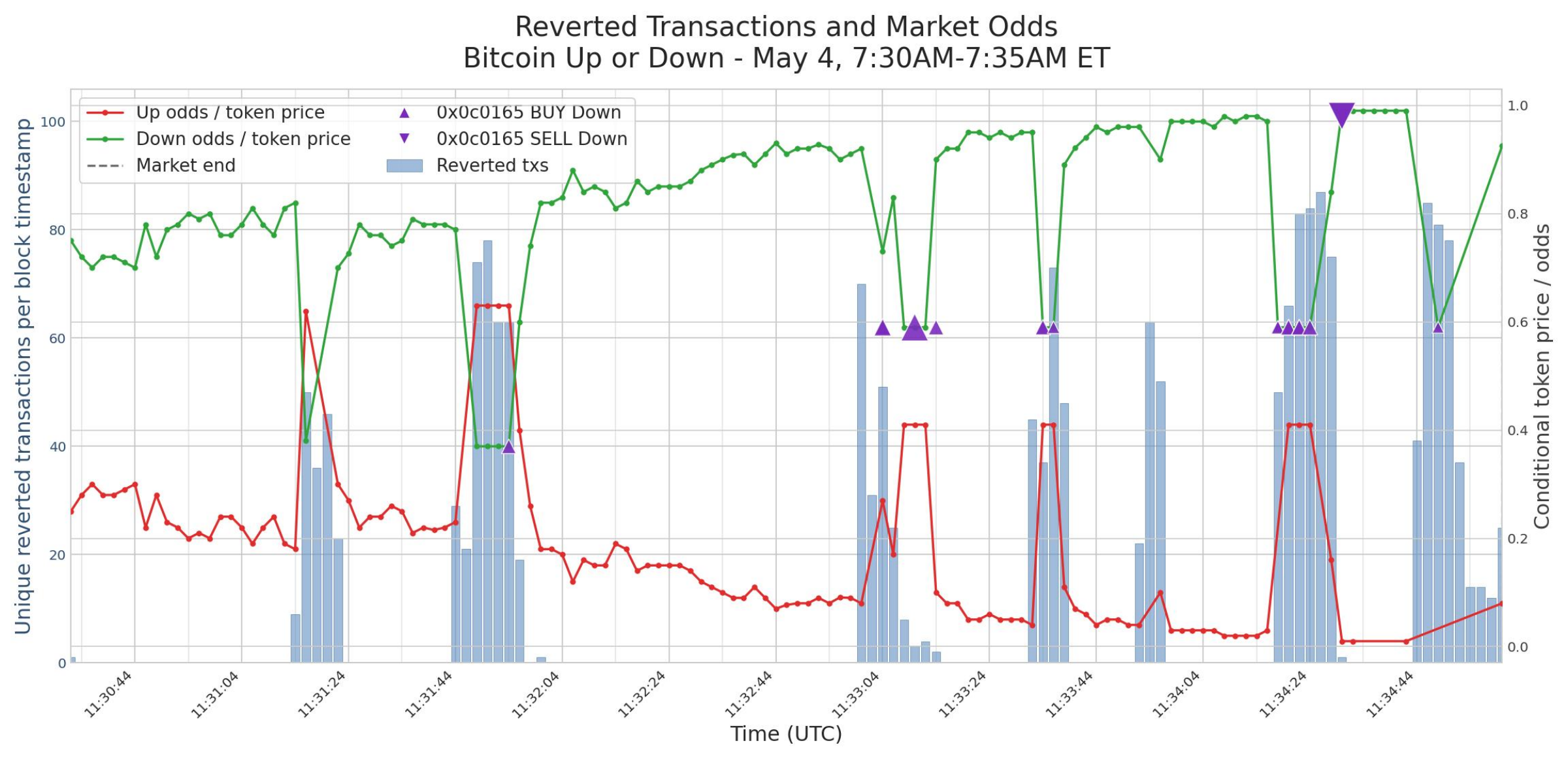}
    \caption{Order-book clearing in the Bitcoin Up and Down 5m market
      (UTC11:30--11:35) on May 4, 2026: per-block reverted settlement counts
      (bars) and volume-weighted \textsc{Up}/\textsc{Down} prices (lines), with
      the highlighted account \texttt{0x0c165\dots} buying \textsc{Down}'s 
      trading history.}
    \label{fig:btc-reverts-odds}
\end{figure}

Case Study~1 concerned reverts near or after market closure. We now turn to a second,
more complicated pattern \emph{in the middle} of an active
market: bursts of reverted settlements that 
clear the order book and
let a single party trade against 
the resulting thin liquidity at favorable
prices.
Figure~\ref{fig:btc-reverts-odds} illustrates 
the pattern using an example of another BTC
Up/Down 5-minute market. The bars count reverted settlement transactions in each
block, while the lines track the volume-weighted price of matched \textsc{Up} and
\textsc{Down} tokens at one block. The figure shows two striking findings. 
First, the reverts form seven sharp
clusters. Within each cluster, a single address is the responsible party 
for almost all of
the reverts: the top-party shares of the seven clusters are 99.4\%, 99.7\%,
44.7\%, 99.5\%, 100.0\%, 99.8\%, and 91.4\%, so 6/7 are over
91\%-dominated by one responsible party. 
This concentration is itself very abnormal. 
Second, each revert cluster coincides with a sharp 
move in the matched price, and
one account, which we label \texttt{0x0c165}, repeatedly profits from it.
Every settled order of this account is a \emph{maker} order. 
As the figure shows,
whenever a revert cluster pushes the price down, this account's resting maker
orders buy \textsc{Down} tokens at the 
very bottom of the cluster window, and it
later sells them near market close. 
It thus consistently acquires shares more
cheaply than other participants, and does so 
precisely when the book has just
been cleared.

The mechanism behind this can be explained as follows. 
Under normal conditions, 
multiple market
makers post maker orders that keep the order book 
deep and the gap between best bid and best ask tight. 
A cluster of reverts removes many resting maker orders from 
the book in a short window, leaving only
a thin set of surviving maker orders  
that may be stale and far from the fair price. 
An incoming taker order is then forced to 
match against these remaining unfavorable
quotes, producing the transient price spike 
visible in the figure. A party that
anticipates the clearing and keeps a maker order can
thus capture
this mispricing before the displaced market makers refill 
liquidity and the
price comes back. In effect, the reverts serve to momentarily 
exclude competing
market makers.

To test whether this pattern generalizes, we 
systematically extracted reverted transaction
clusters across all markets in scope. 
We define a cluster as consecutive
blocks in which each block contains at least $5$ 
reverted transactions of the
same market, merging those whose gap is $\leq 2$ blocks. This yields
$64,604$ clusters across $30,406$ distinct markets. 
For each cluster $[t_0, t_1]$, we
compute the average \textsc{Up} and \textsc{Down} prices over a pre-window
$(t_0\!-\!10,\, t_0)$, the cluster window $(t_0,\, t_1)$, and a post-window
$(t_1,\, t_1\!+\!10)$ (in blocks), and compare the cluster price against its
neighbors.

The example's two findings both recur at scale.
The reverts are overwhelmingly
single-party: in 49,567 of the 64,604
clusters (77\%), one address is responsible
for at least 90\% of the reverted transactions.
10,017 clusters have at least one matched order 
in all three windows: pre, cluster, and post.
The price disruptions are also
real rather than incidental: comparing 
each cluster-window price against its
pre- and post-windows, it lands above 
both neighbors (a price spike) in 3,265
clusters and below both (a price dip) in 3,133, 
whereas it sits smoothly between
them in only 3,619. 
In other words, a genuine spike or dip happened
within 63.88\% of the revert clusters, 
 providing arbitrage opportunities for makers 
who anticipate the clearing. 
Despite we could 
not find clear connection between the 
responsible parties and the makers who profit,
the pattern is consistent with the hypothesis 
that the reverts are a deliberate 
strategy to clear the book and exclude competing 
market makers.
As with avoided losses in Case Study~1, the 
clusters concentrate in crypto
markets (87\%), yet the remaining
categories include: sports (2.39\%), esports (2.00\%),
and politics (1.01\%).

\section{Auditing Procedure and Results}
\label{sec:auditing}

As discussed in Section~\ref{sec:ghost-filled-orders}, 
ghost-filled orders arise from an atomicity gap 
between off-chain matching and on-chain settlement. 
This gap stems from fundamental design choices, 
not specific to a single implementation.
It can appear in any non-custodial blockchain-based 
prediction market that accepts an order off-chain, 
reports the orders as filled/matched, and only later 
attempts to realize the filled/matched through 
an on-chain settlement transaction. However, 
it remains unclear whether the same ghost-filled
order issue can still exist in other 
blockchain-based prediction markets, and if so,
under what circumstances they can be exploited, and 
more importantly, when will a potential attacker be
incentivized enough to exploit them?

We therefore develop a systematic audit methodology 
for testing prediction markets. Our goal is 
not only to determine whether a settlement transaction 
can be made to revert, but also to measure 
how late the attacker can do so after receiving 
an off-chain fill. 
This timing distinction 
is important: a market may be mechanically vulnerable 
if a user can invalidate a settlement precondition, 
but the vulnerability becomes economically exploitable 
only when the user can wait for additional information 
before deciding whether to invalidate the order.

\noindent\textbf{Audit interfaces.}
We consider an attacker with the same capabilities 
as a normal technical trader without any privileged access.
The attacker can submit orders through the public 
market API and can also submit transactions 
directly to the underlying blockchain. 
Such transactions include token transfers, 
approval revocations, 
and market-specific cancellation transactions.
The attacker does not control the market operator, 
or off-chain matching engine.

\begin{property}[No damage to other orders]
\label{prop:no-cross-order-cancellation}
For any order $o$ controlled by user $u$, an invalidation
of $o$ by $u$ (either on-chain or off-chain) must not change
the settlement outcome or
order-book state of any other valid order
$o'$ where $o'$ is not controlled by $u$.
\end{property}

Invalidating one order should not affect the 
behavior of any other order.
Bundled on-chain settlement designs inherently 
violate this property: when orders are
submitted on-chain in matched bundles, invalidating a single order forces the
rest of the bundle to be cancelled as well.
Violating Property~1 means ghost-filled orders 
are \textit{exploitable} in this
prediction market, allowing a malicious user to 
damage other users' orders. Even
when such an attack yields no direct profit, it 
remains a genuine vulnerability,
since one user can interfere with the 
orders of others.

\begin{property}[No asymmetric post-fill invalidation window]
\label{prop:no-asymmetric-window}
Let $o$ be an order submitted by user $u$.
Let $t_{\mathsf{fill}}(o)$ be the first time at 
which $u$ can no longer cancel
$o$ through the public API, for example, when the 
market API reports $o$ as
filled or locked, and let $t_{\mathsf{settle}}(o)$ be the time at which $o$ is
settled on-chain. For any time $t$ such that
$t_{\mathsf{fill}}(o) < t \leq t_{\mathsf{settle}}(o)$,
$u$ should not be able to invalidate $o$ on-chain at time $t$.
\end{property}

This property ensures that a user's on-chain invalidation 
window must be no longer than the
API's cancellation window. 
Violating this property gives users who understand
blockchain and smart contracts an unfair technical 
advantage over API-only
users: they can wait until after blockchain-unaware users cannot 
cancel their orders, 
observe how the market moves, and only
then decide whether to let the order settle, 
exactly the profitable strategy
introduced in Section~\ref{subsec:attack-loss-avoidance}.
Violating Property~2 therefore means ghost-filled 
orders are not only
\textit{exploitable} but also \textit{profitable} in a 
prediction market,
giving a potential attacker a strong economic incentive to act. We regard this
as a critical vulnerability, since the advantage it confers can be exploited by
anyone with basic blockchain knowledge.

\subsection{Audit Methodology}

Our audit has two steps. 
First, we inspect the market's settlement contracts, 
following the same methodology in Section~\ref{subsec:preconditions}, 
and identify user-controlled preconditions that can be 
invalidated before settlement. 
We then test exploitability by submitting a 
small order through the public API and attempting 
to make its later settlement 
transaction revert by invalidating 
one such precondition on chain. A successful 
revert already means
violation of Property~1, which represents 
ghost-filled orders are 
exploitable for this market.
Second, we measure how long an attacker can wait after an off-chain 
fill (the point at which users can no longer cancel 
their orders through the API)~\footnote{We use taker orders 
in our auditing procedure,
because, once submitted, 
a taker order either immediately fills or directly gets killed.
So it is easy to observe the API-reported order fill time.} 
before invalidating settlement on chain. 
Algorithm~\ref{alg:gfo-audit} gives the procedure. 
Let $T$ denote the time at which the API reports 
the order $o$ as filled, and let $T+X$ 
denote the time at which the attacker submits an order-invalidating 
transaction. 
The audit initializes the best successful delay 
$X^\star$ and starts from delay $X=0$ (line~1), 
then iterates over candidate delays (line~2). For each 
candidate delay, it submits a taker order and fills a small order 
through the API, recording both the order $o$ and fill 
time $T$ (line~3). It then waits until $T+X$ (line~4) and 
sends the on-chain invalidation transactions (line~5). 
Then the algorithm observes the settlement outcome $b$,  
If the settlement transaction containing the order $o$ 
reverts, 
the algorithm record $X$ as a successful delay
(lines 7--10). The audit then increases the delay by the step 
size $\delta$ (line~10) and finally returns the 
maximum successful delay $X^\star$ in line~12. 
A positive $X^\star$ indicates a violation 
of Property~2: ghost-filled orders
are not only exploitable but also profitable in this market, because the
attacker gains additional reaction time to carry out the profit-driven
manipulation such as two scenarios described in 
Section~\ref{subsec:attack-loss-avoidance}.

\begin{algorithm}[t]
\caption{Ghost-filled-order audit}
\label{alg:gfo-audit}
\begin{algorithmic}[1]
\Require Market $M$, step size $\delta$, order-invalidating method 
$\mathcal{I}$, maximum delay $\mathsf{MaxDelay}$
\Ensure Maximum successful delay $X^\star$
\State $X^\star \gets \bot$, $X \gets 0$
\While{$X \leq \mathsf{MaxDelay}$}
    \State $(o,T) \gets \mathsf{FillViaAPI}(M)$ 
    \State $\mathsf{Sleep}(X)$
    \State $\mathsf{SendOnChain}(\mathcal{I})$
    \State $b \gets \mathsf{Observe}(o)$
    \If{$b=\mathsf{reverted}$}
        \State $X^\star \gets X$
    \EndIf
    \State $X \gets X + \delta$
\EndWhile
\State \Return $X^\star$
\end{algorithmic}
\end{algorithm}

\subsection{Audit Results}

\noindent\textbf{Markets tested.}
Beyond the Polymarket case study, we audited 
three additional blockchain-based prediction markets 
selected by trading volume among systems whose relevant 
contract addresses were public. 

\begin{table}[tbp]
  \centering
  \caption{Audit results for three additional blockchain-based prediction
    markets.}
  \label{tab:audit-results}
  \footnotesize
  \setlength{\tabcolsep}{6pt}
  \begin{tabular}{@{}lccccl@{}}
    \toprule
    Market & \shortstack[c]{Exploit-\\able?}
      & PoC & \shortstack[c]{Profit-\\able?}  & \shortstack[c]{Monitor\\  \& Ban?}  & Disclosure \\
    \midrule
    Market X$^\dagger$ & \checkmark & \checkmark & \checkmark  & $\times$     & Reported \\
    Market Y$^\dagger$ & \checkmark & \checkmark & $\times$    & $\times$     & Reported \\
    Market Z$^\dagger$ & \checkmark & \checkmark & $\times$    & \checkmark   & Acknowledged \\
    \midrule
    \multicolumn{6}{c}{Combined 7-day volume of \$61.98M$^\dagger$} \\
    \bottomrule
  \end{tabular}
  \begin{minipage}{\linewidth}
  {\footnotesize $^\dagger$\,For responsible disclosure, we temporarily
  withhold the names of the three audited markets and their individual 7-day
  trading volumes until they have patched the issue.}
  \end{minipage}
\end{table}

\noindent\textbf{Findings.}
Table~\ref{tab:audit-results} summarizes our findings.
All three audited markets proved exploitable.
In each market, we successfully submitted a small taker 
order through the public API,
observed that the order had been filled off-chain, and
then invalidated an on-chain precondition
to revert the corresponding settlement transaction.
Moreover, the settlement transactions we observed
were batched with orders from other users.
By invalidating our own order, we could
therefore affect unrelated orders in the same batch.
In our tests, these unrelated orders were not
reliably returned to the order book after the
settlement failure, violating Property~1.
Market Z acknowledged the vulnerability as a 
known gap, which confirms our
findings regarding exploitability.\footnote{Market Z's response, quoted in
  part: ``we can confirm this is a known gap\ldots'' and 
  ``this attack vector is
  quite expensive and impractical\ldots we 
  already have mitigation measures in
  place to detect and prevent this type of 
  activity.'' While we respectfully
  disagree with their assessment of 
  attack cost and practicality, their
  acknowledgement confirms exploitability.}

Only Market X proved both exploitable 
and profitable, violating Property~2.
The situation is particularly severe for Market X. 
Market X is specially designed for sports markets and
introduces an up-to-10 second delay between off-chain 
fill and on-chain settlement
to reduce latency advantages between automated traders and ordinary users.
Despite users can no longer cancel orders through the API during the 
delay period, users still retain on-chain control over the assets and
approvals needed for settlement during this delay.
Consequently, the delay has the opposite security effect:
API-only users see their orders as locked,
while blockchain-aware users can still transfer funds, revoke approvals,
or otherwise invalidate settlement.
The fairness mechanism therefore becomes an attack amplifier,
making the attack not only exploitable but also profitable.

\subsection{Defenses}
\label{subsec:defenses}
We discuss three orthogonal defense mechanisms.

\noindent\textbf{Lock Resources of Filled Orders.}
One defense is to ensure that, once 
an order is filled, the resources needed for settlement 
are no longer under unilateral user control. 
When an order is submitted for matching, 
the system should check and enforce: 
all the preconditions required to settle the order 
must be committed to the settlement transaction, 
and the user should not be able to invalidate any of them
before settlement.
During our audit, we found that Polymarket's v2 deposit 
wallet design follows this 
direction for new users. Deposit wallets are controlled 
by both the user and Polymarket. 
User actions such as order 
cancellation, approval changes, and token transfers 
have to be submitted through 
Polymarket. 
This allows Polymarket to detect and 
prevent precondition invalidations before settlement. 
However, this design also weakens the original 
non-custodial promise, since user operations are submitted entirely 
via Polymarket. 
Moreover, legacy v1 accounts may still retain the ability 
to invalidate settlement preconditions directly.

\noindent\textbf{Operate Own Blockchains to Guarantee Atomicity.}
Several blockchain-based applications, 
such as Hyperliquid~\cite{hyperliquid2026},
operate their own blockchains. 
If prediction markets operate their own
blockchains and design their consensus 
protocols to enforce per-transaction
checks that ensure order matching and 
settlement occur atomically, then both
Property~1 and Property~2 are satisfied. 
This can be achieved either by
introducing a special transaction type for settlement 
transactions or by performing the necessary 
checks before finalizing a block. The atomicity gap
between off-chain matching and on-chain settlement can be 
completely eliminated
by merging all the components into the consensus protocol. 
Alternatively, if prediction markets create and operate
their own layer-2 blockchains~\cite{gudgeon2020sok}, they can redesign the
sequencer to enforce atomicity.

\noindent\textbf{Use Monitoring and Account Banning only as Mitigation.}
Reactive defenses such as detecting precondition 
invalidations and banning accounts
can raise the attacker's cost thus partially mitigate the issue.
Since fresh wallet creation is inexpensive on blockchain 
and a fast 
attacker may complete the exploit before
being detected and banned, a potential attacker can still 
conduct the attack by using a large number of fresh accounts, so this approach is not ideal.
Similarly, ordering-based defenses such as prioritizing settlement transactions
or randomizing settlement schedules can also reduce the attacker's timing control and
lower attack profitability to mitigate the issue. 
During our audit, we found that both Polymarket and Market Z
actively monitor for suspicious precondition invalidation behavior 
and ban accounts accordingly. 

\section{Related Work}
\label{sec:related}

\noindent \textbf{Prediction Market.}
Prediction markets have long been studied as
 mechanisms for aggregating information about future 
 events~\cite{wolfers2004prediction} in economics 
 and finance communities. But only 
 recently we have blockchain-based prediction markets, 
for example, Polymarket~\cite{polymarket} launched in
2020. Recently, some works systematically 
studied the market microstructure and trading behavior on 
Polymarket~\cite{rahman2025sok, dubach2026anatomy}. However,
these works did not touch the security of 
prediction markets. In the prediction market community,
there are independent reports of ghost-filled orders
for different markets on social media~\cite{itslirrato_2026, 
frank2026polymarketOrderAttack}. 
Our work empirically studies
ghost-filled orders at scale and analyzes the underlying
atomicity violation and its financial impact.

\noindent \textbf{Blockchain MEV and Front-Running.}
A large body of prior works have studied adversarial 
and profitable transaction ordering on blockchains:
especially MEV (Maximal Extractable Value) and front-running.
Torres et al. performed a large-scale study of 
front-running on Ethereum~\cite{torres2021frontrunner}.
Zhou et al. studied high-frequency trading and 
sandwich attacks on decentralized on-chain 
exchanges~\cite{zhou2021high}, while Qin et al. 
quantified blockchain extractable value and 
showed that MEV can create consensus-layer 
incentives to fork the chain~\cite{qin2022quantifying}.
Our work shares the same technical grounds 
as these prior works, users can gain profit by 
controlling their transactions position relative 
to other transactions, but in the context of prediction markets. 
As observed in histograms in Table~\ref{tab:stats}, 
many order-invalidating transactions are included 
in the same block as the settlement transaction, 
highly likely resulted from front-running.

\noindent \textbf{TOCTOU and Race Conditions.}
Ghost-filled orders are an instance of the 
classic time-of-check to time-of-use problem.
Traditional TOCTOU work~\cite{bishop1996checking,dean2004fixing,borisov2005fixing} studies races in operating 
systems and file-system APIs, while this work 
studies the same class of atomicity violation in the context of
blockchain-based prediction markets.
Concurrent work by Shen et al.~\cite{shen2026ghosts}, 
released on arXiv
in June 2026 while this paper was under submission, 
investigates 
the same off-chain/on-chain settlement gap. 
While their work focuses on 
realized profits, and cross-chain contract reuse, our work 
emphasizes a precondition-based formalization and 
counterfactual avoided losses and forgone gains. 
While their study observes settlement failures in 
deployed markets, our active audits 
deliberately invalidate matched orders and 
measure the remaining invalidation window 
after API cancellation is disabled, 
distinguishing settlement disruption from an 
economically advantageous post-fill window.

\noindent \textbf{Smart Contract Security.}
A large body of work studies smart contract security
analysis~\cite{tsankov2018securify,schneidewind2020ethor, krupp2018teether, rodler2018sereum}
and transaction tracing~\cite{zhang2020txspector, su2021evil}.
Our work analyzes the smart contracts of prediction markets and adopts similar
techniques to backtrack transactions. Unlike prior work, however, we do not
target vulnerabilities in the smart contract code itself, but rather those
arising from design flaws of blockchain-based prediction market systems.

\section{Conclusion}
\label{sec:conclusion}
This paper presents a systematic
study of ghost-filled orders in 
blockchain-based prediction markets. 
We formalize the problem, 
develop an analysis framework 
that classifies reverted 
settlement transactions and traces
their causes at scale, and show that 
ghost-filled orders can be associated with 
significant financial advantages, 
including over 61M USD in avoided 
losses and frequent arbitrage-like 
opportunities. 
We also introduce a black-box auditing procedure 
and find that multiple deployed 
prediction markets remain vulnerable, 
suggesting that ghost-filled orders are a 
broader design risk that requires stronger 
atomicity guarantees between order 
matching and settlement.

\bibliographystyle{IEEEtran}
\bibliography{refs}

\end{document}